\documentclass[]{aa}
\usepackage{psfig}

\begin{document}

\thesaurus{11.08.1; 11.09.4; 11.17.1; 11.17.4}

\title{Identification of absorbing galaxies towards the
QSO J2233--606 in the Hubble Deep Field South \thanks{ Based on
observations obtained with the NTT at the European Southern
Observatory, La Silla, Chile under programs 61.A--0631 and 62.O--0363;
and with the \hbox{NASA/ESA {\sl Hubble Space Telescope}} by the Space
Telescope Science Institute, which is operated by AURA, Inc., under
NASA contract NAS 5--26555.}}

\author{Laurence Tresse\inst{1,2,8} 
\and Michel Dennefeld$^{1,3}$ 
\and Patrick Petitjean$^{1,4}$ 
\and Stefano Cristiani$^{5,6}$
\and Simon White$^{7}$}

\institute{$^1$ Institut d'Astrophysique de Paris - CNRS, 98bis Boulevard 
Arago, F-75014 Paris, France \\
$^2$ Istituto di Radioastronomia del CNR, Via Gobetti, 101, I-40129 Bologna, Italy \\
$^3$ Universit\'e Pierre et Marie Curie, Paris, France \\
$^4$ DAEC - Observatoire de Paris, 5 Place Jules Janssen, F-92295 Meudon, France \\
$^5$ European Southern Observatory, Karl-Schwarzschild-Str. 2, D-85748 Garching bei M\"unchen, Germany \\
$^6$ Dipartimento di Astronomia, vicolo dell'Osservatorio 5, I-35122 Padova, Italy\\
$^7$ Max-Planck Institut f\"ur Astrophysik, D-85740 Garching, Germany \\
$^8$ Laboratoire d'Astonomie Spatiale, BP8, F-13376 Marseille Cedex 12, France }

\date{}
\offprints{L. Tresse}
\titlerunning{Absorbing galaxies towards J2233--606}
\authorrunning{Tresse et al.}
\maketitle
\markboth{Tresse et al.}{Absorbing galaxies towards J2233--606}

\begin{abstract}

We present spectroscopic observations of galaxies lying within
1\arcmin\ of QSO J2233--606 in the Hubble Deep Field South (HDF-S).
Several are found to be coincident in redshift with absorption-line
systems seen in the HST J2233--606 spectrum.  We detect a new $z_{\rm
em}\!=\!1.336$ QSO with $I\!=\!20.8$ at a projected angular separation of
44.5\arcsec\ (or $\rho\!=\!200h^{-1}$ kpc; $q_0\!=\!0.5$) from
J2233--606. This QSO pair is an ideal target for QSO environment
stu\-dies. Indeed, strong H\,{\small I} Ly$\alpha$ and Ly$\beta$
absorptions are seen at $z_{\rm abs}\!=\!1.3367$ in the J2233--606
spectrum.  The bright spiral galaxy (J2233378--603324), projected at
4.7\arcsec\ from J2233--606, is at $z\!=\!0.570$ (thus $\rho\!=\!18h^{-1}$
kpc). Absorption is seen in the Lyman series at the same redshift but
the weakness of the Lyman break implies $N$(H\,{\small
I})~$\!<\!10^{17}$ cm$^{-2}$.


\end{abstract}

\section{Introduction} 

Recently, a second HST deep field has been observed in the southern
hemisphere (HDF-S; Williams et al. 1999, in preparation). The STIS
field was chosen to contain the QSO J2233--606 ($z_{\rm
em}\!=\!2.238$, $B\!=\!17.5$; Boyle 1997).  Both deep imaging of the
QSO field and spectroscopy of the QSO itself have been obtained
(Gardner et al. 1999, Ferguson et al. 1999). Low- and high-dispersion
spectra from ground-based telescopes are also already available
(Outram et al. 1998, Savaglio 1998, Sealey et al. 1998).  This set of
data offers a unique opportunity to address important questions
related to the connection between galaxies and QSO absorption lines, 
including the absorption cross-section of faint galaxies and the
structure of the inter-galactic medium over the redshift range
1.2--2.2.  Spectroscopy of the brightest galaxies in the full HDF-S,
selected from our $I$-band ground-based image, has been performed
on the ESO-NTT by a collaboration assembled under the auspices of a
European Training and Mobility of Researchers (TMR) network.  Here we
comment on the galaxies closest to J2233--606 and discuss their
possible relation with spectral features of J2233--606.

Sect.~2 briefly presents the target selection, observation and data
reduction, with emphasis on objects close to J2233--606.  In Sect.~3
we analyse their relevance to absorption lines in the J2233--606
spectrum. In Sect.~4 we draw our conclusions.  Throughout the paper,
we adopt $H_0\!=\!100h$ \hbox{km s$^{-1}$ Mpc$^{-1}$} and
$q_0\!=\!0.5$.

\section{Data}

\subsection{Target selection, observation and data reduction}

A 30mn equivalent $I$-band image centered on J2233--606 was obtained
in the ESO Director's discretionary time with the EMMI-NTT red-imaging
channel and made available to us for target selection.  Photometric
calibration has been done with standard stars from Landolt (1992) and
the accuracy of the zero point is better than 0.1 magnitude. Source
extraction and star-galaxy separation were performed with Sextractor
(Bertin \& Arnouts, 1996) resulting in a catalogue of 1159 objects
with $I\!\le\!22.2$ complete at the 90\% level.  The US Naval
Observatory cata\-logue was used as reference for the astrometry. This
gives an accuracy close to 1\arcsec\ for absolute positions and about
three times better for relative positions within the field.
Multi-slit spectroscopy has been carried out with EMMI where about
thirty objects can be observed simultaneously.  We used slits of 1.02
or 1.34\arcsec\ in width, leading to a spectral resolution of
FWHM~$=$~10.6 or 13.9 \AA.  The spectral range lies within 3900--10000
\AA\ but its actual length depends on the location of the object
within the mask.  Reduction was done with standard techniques using an
updated version of Multired under the IRAF reduction package.  The
residuals of the wavelength calibration fits have an r.m.s. of less
than 0.7 \AA, but the positioning of the objects relative to the slit
was accurate to $\sim$0.3\arcsec, leading to wavelength uncertainties
of order 3 \AA.  The accuracy in measuring the wavelengths of lines in
the galaxy spectra is close to one tenth of the resolution element,
i.e. 1. to 1.4 \AA.  Thus the redshift accuracy is mainly limited by
the position of the galaxies within the slits, and our redshifts are
accurate to about 0.001.  Spectra were flux calibrated using standard
stars from Stone \& Baldwin (1984) with no attempt to correct here for
aperture losses. Full details of observations of the complete
sample, data reductions and measured parameters are in Dennefeld et
al. (1999, in preparation).

\subsection{Objects around J2233--606}

\begin{figure}
\vspace{-1.cm}
\centerline{\vbox{\psfig{figure=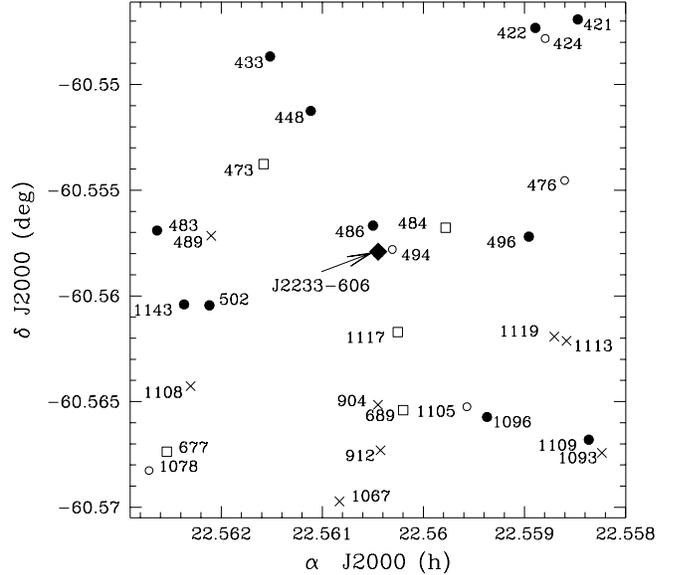,height=9cm,width=9cm,angle=0}}} 
\caption[fig1]{Objects at $I\!\le\!22.2$ around J2233--606 (diamond
symbol) with the running number of our photometric catalogue.  North is
at the top and East to the left. Objects with a compactness of more than
0.95, i.e. likely to be stars, are shown with crosses. For the
remaining objects, circles represent the 16 objects at $I\!\le\!21.5$ of
which 11 spectra have been obtained (filled circles).  Squares
represent fainter objects.
\label{fig1}}
\end{figure}

\begin{figure}
\vspace{-3.0cm}
\centerline{\vbox{\psfig{figure=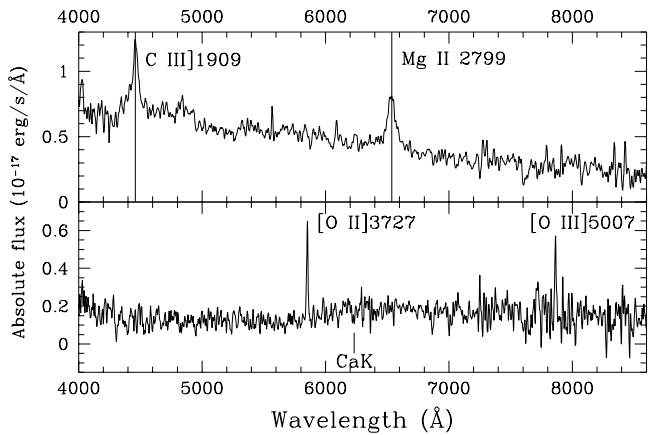,height=9cm,width=9cm,angle=0}}} 
\caption[fig2]{Spectrum of Q433 (top) and G486 (bottom). \label{fig2}}
\end{figure}

Galaxies for which we have performed spectroscopy are shown by a
filled circle in Fig.~\ref{fig1}. Those galaxies lying within
1\arcmin\ of J2233--606 are listed in Table~\ref{tab1} two of which
are of particular interest: Q433 (J2233415--603255) is another QSO,
and G486 (J2233378--603324) is a spiral galaxy at only 4.7\arcsec\
from J2233--606. Their spectra are displayed in Fig.~\ref{fig2}. The
redshift of Q433, $z\!=\!1.336\!\pm\!0.001$, is determined from the
broad Mg\,{\small II} $\lambda$2799 and C\,{\small III}] $\lambda$1909
emission lines.  Also seen are the broad Fe\,{\small II}
$\lambda\lambda$2400, 2600 complex, and the as yet unidentified broad
feature around 2100 \AA\ (see for instance Francis et al. 1991). The
observed (V$-$I) spectral index is 0.5; using the corresponding
spectral energy distribution gives $M(B)\simeq-21.5+5\log h$.  G486
is at $z\!=\!0.570\!\pm\!0.001$ ([O\,{\small II}] $\lambda3727$,
[O\,{\small III}] $\lambda5007$, CaK) and has a late-type spectrum
consistent with its Sc morphology and the presence of numerous
H\,{\small II} regions in the HST image. The observed (V$-$I)
spectral index is 1.5, giving $M(B)\simeq-19.8+5\log h$
(i.e. $L\!\sim\!L^{*}$).

\begin{table}
\caption[tab1]{Galaxies within a 1\arcmin\ radius of J2233-606 with
the running number of our photometric catalogue (ID), the 
coordinates ($\alpha$, $\delta$), the $I$-band apparent magnitude
($I$), the heliocentric redshift ($z$), the angular separation
($\theta$) to J2233--606, and the impact parameter
($\rho/h$).
\label{tab1}}
\begin{tabbing}
\hspace{1.3cm}\=\hspace{1.5cm}\=\hspace{1.8cm}\=\hspace{1cm}\=\hspace{1.5cm}\=\hspace{1cm}\=\hspace{1cm}\kill
$\ \ $ID \> $\alpha$ (J2000) \> $\delta$ (J2000)  \> $\ \ \ I$  \> $\ \ \ z$ \> $\ \ \theta$ \>  $\rho$/$h$ \\
$\ \ $ \> $\ \ \ $[$\degr$] \> $\ \ \ \ $[$\degr$]  \> $\ \ \ $  \> $\ \ \ $ \> \  [\arcsec]  \> [kpc] \\
G422  \> 338.38343 \> $-$60.5473   \> 20.64 \> 0.6465  \> 56.8 \>  220    \\
Q433  \> 338.42273 \> $-$60.5487   \> 20.78 \> 1.3360  \> 44.5 \>  190    \\
G448  \> 338.41668 \> $-$60.5513   \> 19.55 \> 0.5800  \> 29.8 \>  112    \\
G483  \> 338.43954 \> $-$60.5569   \> 19.91 \> 0.3302  \> 58.4 \>  170    \\
G486  \> 338.40752 \> $-$60.5567   \> 20.32 \> 0.5704  \> $\ \,4.7$ \>  $\ \:18$    \\
G496  \> 338.38436 \> $-$60.5572   \> 21.48 \> 0.4148  \> 39.9 \>  130    \\
G502  \> 338.43177 \> $-$60.5604   \> 18.57 \> 0.2268  \> 45.0 \>  103    \\
G1096 \> 338.39056 \> $-$60.5657   \> 20.79 \> 0.4147  \> 40.1 \>  130    \\
G1143 \> 338.43553 \> $-$60.5604   \> 21.29 \> 0.066?  \> 51.5 \>  $\ \:45$    
\end{tabbing}
\end{table} 

\section{Absorption lines in J2233--606}

\subsection{Q433 at $z_{\rm em} = 1.336$}

\begin{figure}
\centerline{\vbox{
\vspace{-8.8cm}
\psfig{figure=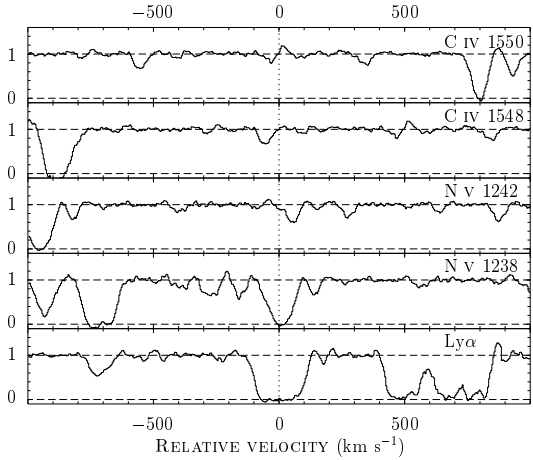,height=14.8cm}}} 
\caption[fig3]{Absorptions for the $z_{\rm abs}\!=\!1.3367$ absorption-line 
system in the normalized J2233--606 spectrum. 
The Ly$\alpha$ line profile suggests a multi-component
structure. From bottom to top, H\,{\sc i} Ly$\alpha$,
N\,{\sc v} 1238, N\,{\sc v} 1242, C\,{\sc iv} 1548, C\,{\sc 
iv} 1550. \label{fig3}}
\end{figure}

We searched the HST spectrum for absorptions around $z\!=\!1.336$.
The wavelength ranges of H\,{\small I} Ly$\alpha$ together with
C\,{\small IV} $\lambda\lambda$1548, 1550 and N\,{\small V}
$\lambda\lambda$1238, 1242 around this redshift are shown in
Fig.~\ref{fig3} on a relative velocity scale, $v$.  Strong H\,{\small
I} Ly$\alpha$ and Ly$\beta$ absorption lines are detected at $z_{\rm
abs}\!=\!1.3367$. 
The Ly$\beta$ line however is redshifted in a region of poor S/N 
below the Lyman break of the modera\-tely optically thick system
at $z\!\sim\!1.9$ and is most certainly blended with Ly$\alpha$
absorption at a different redshift.
More than one component are probably present since the con\-ti\-nu\-um level
at the bottom of Ly$\alpha$ goes to zero over about 150 km s$^{-1}$
but neither damping wings nor an associated Lyman break are present.
The total equivalent width of the Ly$\alpha$ line, $W_{\rm
r}\!=\!0.87$ \AA, suggests a neutral hydrogen column density of at
least $N($H\,{\small I}$)\!\sim\!10^{16}$ cm$^{-2}$.  A one-component
fit gives $N($H\,{\small I}$)\!\sim\!5\:10^{15}$ cm$^{-2}$ and a
Doppler parameter $b\!\sim\!50$ km s$^{-1}$. The latter large value of
$b$ provides additional evidence for multiple structure.  We
tentatively fit the line with three components at $v\!\sim\!-51$,
$+17$ and $+87$ km s$^{-1}$ with $N($H\,{\small
I}$)\!\sim\!2.5\,10^{14}$, $1.3\,10^{15}$, $5\,10^{13}$ cm$^{-2}$ and
$b\!\sim\!38$, 33 and 25 km s$^{-1}$ res\-pectively.  There might be a
C\,{\small IV} $\lambda$1548 component at $v\sim-60$ km s$^{-1}$ but
with no obvious C\,{\small IV} $\lambda$1550 counterpart; the latter
could be below the detection limit. N\,{\small V} $\lambda$1242
absorption could be present at $v\!\sim\!+55 $ km s$^{-1}$. The
N\,{\small V} $\lambda$1238 counterpart is unseen because it is
blended with a strong Ly$\alpha$ line; and the associated C\,{\small
IV} absorption is not detected.  An absorption line is seen at the
expected position of O\,{\small VI}~$\lambda$1031 but with no obvious
O\,{\small VI} $\lambda$1037 counterpart; the correspon\-ding part of
the spectrum has a poor S/N ratio however.  The presence of metals in
the cloud is thus questionable; better data in the optical range
will help decide this issue.

The good correspondence between the redshift of Q433 and the Lyman
absorption redshift in the J2233--606 spectrum ($\Delta
z\!\sim\!0.0007\pm0.001$, $\Delta v\!\sim\!90\pm130$ km s$^{-1}$)
might only be coincidence. The absorption could in fact be due to gas
associated with an object in the QSO's immediate en\-vironment.  We note
that the number density of Ly$\alpha$ lines with $N($H\,{\small
I}$)\!>\!10^{16}$ cm$^{-2}$ is about 5 per unit redshift (Petitjean et
al. 1993). Assuming no dependence on redshift, the probability that a
randomly placed Ly$\alpha$ cloud with $N($H\,{\small I}$)\!>\!10^{16}$
cm$^{-2}$ is observed within 200 km s$^{-1}$ from the redshift of Q433
along the line of sight to J2233--606 is smaller than 0.01. This
probability is not highly significant since it is an a-posteriori
statistical argument.  Note that Savaglio et al. (1999) have shown
that the region spanning $z\!\sim\!1.383$--1.460 has a low density of
absorption lines with five lines detected when 16 are expected from
the average density along the line of sight. This possible `transverse
proximity effect' is at odds with the presence of the strong line at
the same redshift as Q433. A similar situation has been observed along
the lines of sight to Q1026--0045A,B, two QSOs at $z_{\rm
em}\!=\!1.438$ and 1.520 respec\-tively, with an angular separation of
36\arcsec, corresponding to an impact parameter of $\sim\!150h^{-1}$
kpc (Petitjean et al. 1998).  A metal-poor associated system is seen
at $z_{\rm abs}\!=\!1.4420\!\sim\!z_{\rm em}^{\rm A}$ along the line
of sight to A, with a complex velocity profile. A strong Ly$\alpha$
absorption is detected along the line of sight to B, redshifted by
only 300 km s$^{-1}$ relative to the associated system in A.

Follow-up spectroscopic studies of the field will investigate whether
this QSO/absorption association is a consequence of the presence of a
gaseous disk, halo or other gaseous structure of radius larger than
200$h^{-1}$ kpc around Q433 or is due to a galaxy at a similar
redshift to Q433.

\subsection {G486 at $z_{\rm em}  = 0.570$}

\begin{figure}
\centerline{\vbox{
\vspace{-8.8cm}
\psfig{figure=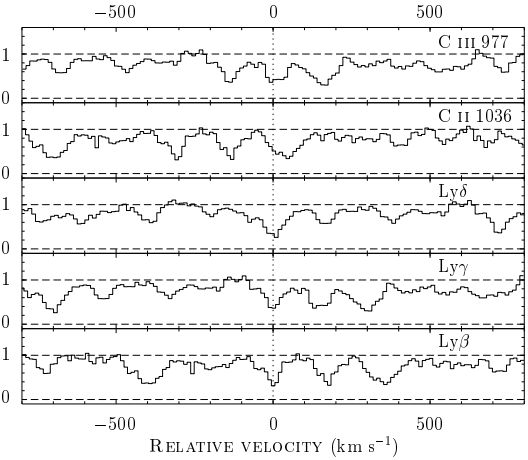,height=14.8cm}}} 
\caption[fig4]{Absorptions for the $z_{\rm abs}\!=\!0.5705$ absorption-line 
system in the normalized J2233--606 spectrum. From bottom to top, H\,{\sc i}
Ly$\beta$, Ly$\gamma$, Ly$\delta$ and C\,{\sc ii} 1036, C\,{\sc iii} 977.
\label{fig4}}
\end{figure}

The line of sight to J2233--606 passes through the disk (seen
approximatively face-on) of a late-type spiral galaxy at an impact
parameter of only $\sim\!18h^{-1}$ kpc. This is a si\-tuation where
conspicuous metal absorptions, and perhaps damped H\,{\small I}
Ly$\alpha$, are expected. H\,{\small I} absorption associated with
this galaxy is seen in the Lyman series (see Fig.~\ref{fig4}) at
$z_{\rm abs}\!=\!0.570\!\pm\!0.01$.  Uncertainties are too large to
reliably estimate the column density from fitting the lines. However,
the fit of the Lyman limit (912\AA) gives $N($H\,{\small
I}$)\!\sim\!10^{16.8}$ cm$^{-2}$ (Outram, private communication).
Because of the poor spectral resolution of the G140L spectrum, the
presence of C\,{\small III} $\lambda$977 and C\,{\small II}
$\lambda$1036 cannot be ruled out, and the C\,{\small IV} and
Al\,{\small III} doublets are most certainly blended. There is no
Fe\,{\small II} $\lambda$2600 absorption at $\lambda$4083.3 in the AAT
spectrum (Outram et al. 1998) down to a 3$\sigma$ limit $W_{\rm
r}\!=\!0.03$ \AA.  The lack of Fe\,{\small II} absorption is
consistent with a low H\,{\small I} column density. Note that
Fe\,{\small II} $\lambda$2382 is lost in a strong Ly$\alpha$ complex.

It is well established that bright ($L_{\rm K}\!>\!0.1~L^{*}_{\rm
K}$) galaxies within 40$h^{-1}$ kpc from the line of sight to a QSO
produce strong ($W_{\rm r}\!>\!0.3$ \AA) Mg\,{\small II} absorption
(e.g. Bergeron \& Boiss\'e 1991, Steidel et al. 1994) whereas fainter
galaxies with a similar range of impact parameters do not produce
detectable metal-line absorptions (Steidel et al. 1997). In the
present case, a weak absorption line at $\lambda$4390.66 is detected
both in the AAT spectrum and in a spectrum recently obtained at ESO
(V. D'Odorico et al., private communication). In the ESO spectrum,
$W_{\rm obs}\!=\!0.18\!\pm\!0.04$ \AA\ is observed. This line is probably
Mg\,{\small II} $\lambda$2796 at $z\!=\!0.5701$.  The limit on the
corresponding weaker Mg\,{\small II} $\lambda$2803 line is consistent
with the optically thin case.  The Mg\,{\small II} absorption is quite
weak for a  galaxy with $L\!\sim\!L^{*}$ and such a small impact parameter: 
this is inconsistent with the correlation between the impact parameter 
and the strength of the absorption claimed by Lanzetta \& Bowen (1990).

\subsection{Other galaxies around J2233--606}

A single component Mg\,{\small II} system is seen at $z_{\rm
abs}\!=\!0.4143$  $b\!=\!7$ km s$^{-1}$ and
$\log N($Mg~{\small II}$)\!=\!12.8$ (Outram et al. 1998).  We observe two
galaxies at a distance smaller than 40\arcsec\ (or 130$h^{-1}$ kpc)
from J2233--606 at redshifts $z_{\rm em}\!\sim\!0.4147$ and 0.4148, (G1096
and G496 in Table~\ref{tab1}), while a third one with $z_{\rm
em}\!\sim\!0.4147$ (G1109) is slightly outside the 1\arcmin\ radius
($\theta\!=\!63$\arcsec).  The well-defined $I$-band selected CFRS
redshift distribution gives 0.38$\pm$0.02 and 0.52$\pm$0.04 gala\-xies
at $I\le22$ by square arcmin in the respective redshift ranges
[0.30--0.40] and [0.40--0.50] (see Lilly et al. 1995). Thus the three
galaxies observed in a 0.0001 redshift range represent a density far
in excess of the expected mean.  This overdensity of galaxies at
$z\!\sim\!0.4147$ suggests that other objects closer to the QSO are
responsible for the absorption. A possible candidate is object G484
(see Fig.~\ref{fig1}), at a distance of 18.2\arcsec, resolved in the
HST image into an interacting pair of spirals.

For the other galaxies (G1143, G502, G483), no conspicuous Mg\,{\small
II} is found, i.e. the $W_{\rm r}$ limit at 3$\sigma$ is 0.10, 0.13 and 0.05
\AA\ respectively at $z_{\rm abs}\!=\!0.066$, 0.227, 0.330. This is
consistent with the halo radius-luminosity scaling-law found for
Mg\,{\small II} absorption-selected galaxies (Bergeron \& Boiss\'e
1991, Steidel et al. 1994).

\section{Conclusions} 

We have carried out spectroscopic observations of galaxies around 
J2233--606 and searched for associated absorption-line systems 
in its spectrum. We find the following. 
\begin{description}
\item[(i)] Q433 (J2233415--603255; $z\!=\!1.336$,
$M(B)\!\simeq\!-21.5\!+\!5\log h$, $\rho\!=\!190h^{-1}$ kpc) is an ideal
target for QSO en\-vi\-ron\-ment studies. Strong H\,{\small I} Ly$\alpha$
and Ly$\beta$ absorpti\-ons are seen at $z_{\rm abs}\!=\!1.3367$ in the
spectrum of J2233--606.  The good redshift agreement might be
coinciden\-ce; however a similar case has already been obser\-ved by
Petitjean et al. (1998). Further investigations are nee\-ded to decide
whether a group of galaxies associated with Q433 or a structure of
radius larger than 200$h^{-1}$ kpc around Q433 is responsible for the
H\,{\small I} absorption.
\item[(ii)] The nearly face-on Sc spiral (J2233378--603324;
$z\!=\!0.570$, $M(B)\!\simeq\!-19.8\!+\!5\log h$, $\rho\!\sim\!18h^{-1}$
kpc) is an excellent target for associated absorption-line system
studies. We find absorption in the Lyman series. The associated Lyman
limit, and the absence of Fe\,{\small II} in the J2233--606 spectrum
are consistent with a H\,{\small I} column density of less than
10$^{17}$ cm$^{-2}$. The Mg\,{\small II} absorption is weak for a
galaxy with $L\!\sim\!L^{*}$ and such a small impact parameter; this
is inconsistent with a tight correlation between impact parameter and
absorption strength.
\item[(iii)] A Mg\,{\small II} absorption system is detected at
$z_{\rm abs}\!=\!0.4143$ in the J2233--606 spectrum. At the same time, there
is an overdensity of galaxies at $z\!\sim\!0.4147$, so other
objects closer than 40\arcsec\ to J2233-606 are expected at the same
redshift.  G484 ($I\!=\!21.5$) is a good candidate for being related to
this absorption.
\end{description}

\acknowledgements{We thank P. Outram for his comments. This program
has been conducted with partial support by the Training and Mobility
of Researchers network under contract FMRX-CT96-0086: 'The
Formation and Evolution of Gala\-xies'. LT acknowledges financial
support by the same network.}

\end{document}